\documentclass[pre,aps,preprint,showpacs,preprintnumbers,amsmath,amssymb,secnumroman,eqsecnum]{revtex4}

\usepackage{graphicx}
\usepackage{dcolumn}
\usepackage{bm}

\newcommand{\beq}{\begin{equation}}
\newcommand{\eeq}{\end{equation}}
\newcommand{\beqa}{\begin{eqnarray}}
\newcommand{\eeqa}{\end{eqnarray}}
\newcommand{\vc}[1]{\mbox{\boldmath $#1$}}

\newcommand{\vol}[1]{{\bf #1}}

\newcommand{\du}[1]{{\bf\sf #1}}

\begin{document}


\title{Optimal translational swimming of a sphere at low Reynolds number}

\author{B. U. Felderhof}

 \email{ufelder@physik.rwth-aachen.de}
\affiliation{Institut f\"ur Theorie der Statistischen Physik\\ RWTH Aachen University\\
Templergraben 55\\52056 Aachen\\ Germany\\
}%

\author{R. B. Jones}

 \email{r.b.jones@qmul.ac.uk}
\affiliation{Queen Mary University of London, The School of
Physics and Astronomy, Mile End Road, London E1 4NS, UK\\}%

\date{\today}

\begin{abstract}
Swimming velocity and rate of dissipation of a sphere with surface distortions are discussed on the basis of the Stokes equations of low Reynolds number hydrodynamics. At first the surface distortions are assumed to cause an irrotational axisymmetric flow pattern. The efficiency of swimming is optimized within this class of flows. Subsequently more general axisymmetric polar flows with vorticity are considered. This leads to a considerably higher maximum efficiency. An additional measure of swimming performance is proposed based on the energy consumption for given amplitude of stroke.
\end{abstract}

\pacs{47.15.G-, 47.63.mf, 47.63.Gd, 87.17.Jj}
\maketitle
\section{\label{I}Introduction}

The subtlety of the theory of swimming at low Reynolds number has not always been fully appreciated. It is important to have simple examples for which calculations can be performed in detail. The first such example was furnished by Taylor \cite{1} in his seminal work on the swimming of an undulating planar sheet immersed in a viscous incompressible fluid. Soon after,
Lighthill \cite{2} studied the swimming of a sphere. He considered a squirming sphere with surface displacements in the spherical surface. His work was extended by Blake \cite{3}, who considered the full class of surface displacements.

 The goal of the theory is to calculate the swimming velocity and the rate of dissipation in the fluid for given time-periodic deformations of the body. The rate of dissipation equals the power necessary to achieve the swimming motion. Shapere and Wilczek formulated the problem in terms of a gauge field on the space of shapes \cite{4}. They pointed out \cite{5} that the measure of efficiency of a stroke introduced by Lighthill and Blake is not appropriate. In low Reynolds number swimming, unlike in the problem of Stokes friction, the power is proportional to the speed, rather than the square of the speed. As a measure of efficiency Shapere and Wilczek therefore introduced a dimensionless number measuring the ratio of speed and power, rather than the ratio of speed squared and power.

The theory of swimming at low Reynolds number is based on the Stokes equations \cite{6}. In earlier work we have extended the theory to include the rate of change of fluid momentum, as given by the linearized Navier-Stokes equations \cite{7}. As an example we studied small-amplitude swimming of a deformable sphere \cite{8}, and found the optimum efficiency for the class of swimming motions for which the first order flow velocity is irrotational. Our definition of efficiency was analogous to that of Shapere and Wilczek.

The calculation based on the linearized Navier-Stokes equations was rather elaborate. It turns out that for irrotational flow the inertial effect vanishes, so that for this class of fluid motions it suffices to use the Stokes equations. This allows a simpler formalism and easier calculations. In the following we discuss the theory on the basis of the Stokes equations, and in addition derive some new results. The stroke of maximum efficiency involves a significant contribution of high order multipoles. This leads us to consider an additional measure of swimming performance, allowing minimization of the energy consumption at fixed amplitude of stroke. We provide a numerical estimate of speed and power for optimal swimming via potential flow of a typical bacterium. Customarily the speed is calculated for given power from Stokes drag \cite{9}.

 In the first part of the article we restrict attention to axisymmetric irrotational flow. The fluid flow velocity can be derived from a scalar potential which satisfies Laplace's equations. It is therefore natural to introduce multipoles in analogy to electrostatics \cite{10}. To linear order the pressure disturbance vanishes. The swimming speed and the power are bilinear in the surface displacements. The class of potential flows is important because of the connection to inviscid flow theory based on the full set of Navier-Stokes equations, as relevant for swimming at high Reynolds number \cite{11}.

 Subsequently we study more general axisymmetric polar flow. This involves modes with vorticity and a non-vanishing pressure disturbance, and requires the use of an additional set of multipoles. It turns out that the more complicated flow with vorticity leads to a significantly higher maximum efficiency than found for potential flow. Again we consider the measure of swimming performance based on energy consumption at fixed amplitude, and provide a numerical estimate for a typical bacterium.

\section{\label{II}Flow equations}

We consider a flexible sphere of radius $a$ immersed in a viscous
incompressible fluid of shear viscosity $\eta$. At low Reynolds
number and on a slow time scale the flow velocity
$\vc{v}(\vc{r},t)$ and the pressure $p(\vc{r},t)$ satisfy the
Stokes equations
\begin{equation}
\label{2.1}\eta\nabla^2\vc{v}-\nabla p=0,\qquad\nabla\cdot\vc{v}=0.
\end{equation}
The fluid is set in motion by time-dependent distortions of the
sphere. We shall study periodic distortions which lead to swimming
motion of the sphere. The surface displacement
$\vc{\xi}(\vc{s},t)$ is defined as the vector distance
\begin{equation}
\label{2.2}\vc{\xi}=\vc{s}'-\vc{s}
\end{equation}
of a point $\vc{s}'$ on the displaced surface $S(t)$ from the
point $\vc{s}$ on the sphere with surface $S_0$. The fluid
velocity $\vc{v}(\vc{r},t)$ is required to satisfy
\begin{equation}
\label{2.3}\vc{v}(\vc{s}+\vc{\xi}(\vc{s},t))=\frac{\partial\vc{\xi}(\vc{s},t)}{\partial t}.
\end{equation}
This amounts to a no-slip boundary condition. The instantaneous translational swimming velocity $\vc{U}(t)$,
the rotational swimming velocity $\vc{\Omega}(t)$, and the flow pattern $(\vc{v},p)$ follow from the condition that no net
force or torque is exerted on the fluid. We evaluate these quantities by a perturbation expansion in powers of the
displacement $\vc{\xi}$.

In the first part of the article we restrict attention to motions for which to first order in the
displacement the flow is irrotational, so that the flow velocity
is the gradient of a scalar potential,
\begin{equation}
\label{2.4}\vc{v}_1=\nabla\phi_1.
\end{equation}
We specify the surface displacement by assuming an expression for
the first order potential. We assume the flow to be symmetric
about the $z$ axis, so that in spherical coordinates
$(r,\theta,\varphi)$, defined with respect to the center of the
sphere in the rest system, the potential takes the form
$\phi_1(r,\theta,t)$. The potential $\phi_1(r,\theta,t)$ tends to
zero at infinity, and can be expressed as the Poisson integral
\begin{equation}
\label{2.5}\phi_1(r,\theta,t)=\int_{r'<a}\frac{1}{|\vc{r}-\vc{r}'|}\;\rho(r',\theta',t)\;d\vc{r}',
\end{equation}
with a source density $\rho(r,\theta,t)$ localized within the sphere of radius $a$. To first order the pressure remains
constant and equal to the ambient pressure $p_0$. We regard the  source density $\rho(r,\theta,t)$ as given, and define the
surface displacement from
\begin{equation}
\label{2.6}\frac{\partial\vc{\xi}}{\partial t}=\nabla\phi_1\big|_{r=a}.
\end{equation}
Instead of $\vc{\xi}$ we regard $\rho$ as the expansion parameter.
For given source density $\rho(r,\theta,t)$ one can evaluate the
first order potential $\phi_1(r,\theta,t)$ by use of Eq. (2.5).
Hence one finds the first order flow velocity
$\vc{v}_1(r,\theta,t)$ by use of Eq. (2.4). Since this tends to
zero faster than $1/r$, the force exerted on the fluid and the
swimming velocity $U(t)$ vanish to first order. The rotational
velocity $\Omega(t)$ and the torque vanish automatically by
symmetry.

We consider in particular harmonic time variation at frequency
$\omega$, with source density
\begin{equation}
\label{2.7}\rho(r,\theta,t)=\rho_c(r,\theta)\cos\omega
t+\rho_s(r,\theta)\sin\omega t,
\end{equation}
with suitably chosen functions $\rho_s(r,\theta)$ and
$\rho_c(r,\theta)$. Since the no-slip condition is nonlinear, the
solution of the flow problem involves harmonics with all integer
multiples of $\omega$.

We perform a perturbation expansion in powers of the two-component
source density
$\vc{\rho}(\vc{r})=(\rho_c(\vc{r}),\rho_s(\vc{r}))$. To second
order in $\vc{\rho}$ the flow velocity and the swimming velocity
take the form
\begin{equation}
\label{2.8}\vc{v}(\vc{r},t)=\vc{v}_1(\vc{r},t)+\vc{v}_2(\vc{r},t)+...,\qquad
U(t)=U_2(t)+....
\end{equation}
Both $\vc{v}_1$ and $\vc{\xi}$ vary harmonically with frequency
$\omega$, and can be expressed as
 \begin{eqnarray}
\label{2.9}\vc{v}_1(\vc{r},t)&=&\vc{v}_{1c}(\vc{r})\cos\omega t+\vc{v}_{1s}(\vc{r})\sin\omega t,\nonumber\\
\vc{\xi}(\vc{s},t)&=&\vc{\xi}_{c}(\vc{s})\cos\omega
t+\vc{\xi}_{s}(\vc{s})\sin\omega t.
\end{eqnarray}
Expanding the no-slip condition Eq. (2.3) to second order we find
for the flow velocity at the surface
\begin{eqnarray}
\label{2.10}\vc{u}_{1S}(\theta,t)&=&\vc{v}_1\big|_{r=a}=\frac{\partial\vc{\xi}(\theta,t)}{\partial t},\nonumber\\
\vc{u}_{2S}(\theta,t)&=&\vc{v}_2\big|_{r=a}=-\vc{\xi}\cdot\nabla\vc{v}_1\big|_{r=a}.
\end{eqnarray}
Hence the swimming velocity can be evaluated as \cite{7}
\begin{equation}
\label{2.11}
U_2(t)=-\frac{1}{4\pi}\int\vc{u}_{2S}\cdot\vc{e}_z\;d\Omega.
\end{equation}
The time-averaged swimming velocity is given by
\begin{equation}
\label{2.12}\overline{U_2}=-\frac{1}{4\pi}\int\overline{\vc{u}}_{2S}\cdot\vc{e}_z\;d\Omega,
\end{equation}
where the overhead bar indicates a time-average over a period $T=2\pi/\omega$.
The remainder $U_2(t)-\overline{U_2}$ oscillates at frequency $2\omega$.

To second order the rate of dissipation $\mathcal{D}_2(t)$ is
determined entirely by the first order solution. It may be
expressed as a surface integral \cite{7}
\begin{equation}
\label{2.13}\mathcal{D}_2=-2\eta\int_{r=a}\nabla\phi_{1}.(\nabla\nabla\phi_{1}).\vc{e}_r\;dS.
\end{equation}
The rate of dissipation is positive and oscillates in time about a
mean value. The mean rate of dissipation equals the power
necessary to generate the motion.

\section{\label{III}Multipole modulation}

In explicit calculations we expand the source density and the first order potential in spherical
harmonics. We define the solid spherical harmonics $\Phi_l^\pm$ as
\begin{equation}
\label{3.1}\Phi^+_l(r,\theta)=r^lP_l(\cos\theta),\qquad\Phi^-_l(r,\theta)=r^{-l-1}P_l(\cos\theta),
\end{equation}
with Legendre polynomials $P_l$ in the notation of Edmonds \cite{12}. The source density $\rho_l=\Phi^+_l$
inside the sphere generates  a potential proportional to $\Phi^-_l$ outside the sphere. It is natural
to extend the potential and the corresponding velocity field inside the sphere. The first order potential
outside the sphere is expanded as
\begin{equation}
\label{3.2}\phi_1(r,\theta)=\omega a^2\sum^\infty_{l=0}\mu_l\bigg(\frac{a}{r}\bigg)^{l+1}P_l(\cos\theta),\qquad r>a,
\end{equation}
with dimensionless multipole coefficients $\{\mu_l\}$. The corresponding first order potential
inside the sphere is given by
\begin{equation}
\label{3.3}\phi_1(r,\theta)=\frac{1}{2}\omega a^2\sum^\infty_{l=0}\mu_l\bigg[(2l+3)\bigg(\frac{r}{a}\bigg)^l
-(2l+1)\bigg(\frac{r}{a}\bigg)^{l+2}\bigg]P_l(\cos\theta),\qquad r<a.
\end{equation}
This has been constructed such that the potential and its radial derivative are continuous at $r=a$. The
corresponding source density is
\begin{equation}
\label{3.4}\rho(r,\theta)=\frac{\omega}{4\pi}\sum^\infty_{l=0}\mu_l(2l+1)(2l+3)\bigg(\frac{r}{a}\bigg)^lP_l(\cos\theta),
\qquad r<a.
\end{equation}
The first order flow outside the sphere is
\begin{equation}
\label{3.5}\vc{v}_1(r,\theta)=-\omega a\sum^\infty_{l=0}\mu_l\vc{u}_l(r,\theta),\qquad r>a,
\end{equation}
with component field
\begin{equation}
\label{3.6}\vc{u}_l(r,\theta)=\bigg(\frac{a}{r}\bigg)^{l+2}\big[(l+1)P_l(\cos\theta)\vc{e}_r
+P^1_l(\cos\theta)\vc{e}_\theta\big],
\end{equation}
with associated Legendre function of the first kind $P^1_l(\cos\theta)$, in the notation of Edmonds \cite{12}. We note that
\begin{equation}
\label{3.7}\vc{u}_l(r,\theta)=-a^{l+2}\nabla\Phi^-_l(r,\theta).
\end{equation}

For the time-dependent source density of the form Eq. (2.7) the
multipole coefficients are time-dependent and can be expressed as
\begin{equation}
\label{3.8}\mu_l(t)=\mu_{lc}\cos\omega t+\mu_{ls}\sin\omega t.
\end{equation}
These generate the first order flow
\begin{equation}
\label{3.9}\vc{v}_1(r,\theta,t)=-\omega a\sum^\infty_{l=0}\mu_l(t)\vc{u}_l(r,\theta),\qquad r>a.
\end{equation}
The corresponding displacement is
\begin{equation}
\label{3.10}\vc{\xi}(\theta,t)=a\sum^\infty_{l=0}\big[\mu_{ls}\cos\omega
t-\mu_{lc}\sin\omega t\big]\vc{u}_l(a,\theta).
\end{equation}

In the calculation of the mean swimming velocity, as given by Eq.
(2.12), we use the identity
\begin{equation}
\label{3.11}\int_{r=a}(\nabla\Phi^-_k)\cdot(\nabla\nabla\Phi^-_l)\cdot\vc{e}_z\;dS=-4\pi
k(k+1)a^{-2k-2}\delta_{k,l+1}.
\end{equation}
This shows that the mean swimming velocity is given by a sum of products of adjacent multipole coefficients,
\begin{equation}
\label{3.12}\overline{U_2}=\frac{1}{2}\omega
a\sum^\infty_{l=0}(l+1)(l+2)\big[\mu_{lc}\mu_{l+1,s}-\mu_{ls}\mu_{l+1,c}\big].
\end{equation}
We define the multipole moment vector $\vc{\mu}$ as the one-dimensional array
\begin{equation}
\label{3.13}\vc{\mu}=(\mu_{0s},\mu_{0c},\mu_{1s},\mu_{1c},....).
\end{equation}
Then $\overline{U_2}$ can be expressed as
\begin{equation}
\label{3.14}\overline{U_2}=\frac{1}{2}\omega
a(\vc{\mu},\du{B}\vc{\mu}),
\end{equation}
with a dimensionless symmetric matrix $\du{B}$. The upper
left-hand corner of the matrix $\du{B}$, truncated at $l=3$, reads
\begin{equation}
\label{3.15}\du{B}_{03}=\left(\begin{array}{cccccccc}0&0&0&-1&0&0&0&0
\\0&0&1&0&0&0&0&0
\\0&1&0&0&0&-3&0&0
\\-1&0&0&0&3&0&0&0
\\0&0&0&3&0&0&0&-6
\\0&0&-3&0&0&0&6&0
\\0&0&0&0&0&6&0&0
\\0&0&0&0&-6&0&0&0\end{array}\right).
\end{equation}
On the cross-diagonals the numbers $\frac{1}{2}(l+1)(l+2)$ appear for $l=0,1,2,...$.

In the calculation of the rate of dissipation, as given by Eq. (2.18), we use the identity
\begin{equation}
\label{3.16}\int_{r=a}(\nabla\Phi^-_k)\cdot(\nabla\nabla\Phi^-_l)\cdot\vc{e}_r\;dS=-4\pi
(k+1)(k+2)a^{-2k-3}\delta_{k,l}.
\end{equation}
Hence the time-averaged rate of dissipation is given by
\begin{equation}
\label{3.17}\overline{\mathcal{D}_2}=4\pi\eta\omega^2a^3\sum^\infty_{l=0}(l+1)(l+2)(\mu_{lc}^2+\mu_{ls}^2).
\end{equation}
This can be expressed as
\begin{equation}
\label{3.18}\overline{\mathcal{D}_2}=8\pi\eta\omega^2a^3(\vc{\mu},\du{A}\vc{\mu}),
\end{equation}
with a dimensionless diagonal matrix $\du{A}$. The upper left-hand
corner of the matrix $\du{A}$, truncated at $l=3$, reads
\begin{equation}
\label{3.19}\du{A}_{03}=\left(\begin{array}{cccccccc}1&0&0&0&0&0&0&0
\\0&1&0&0&0&0&0&0
\\0&0&3&0&0&0&0&0
\\0&0&0&3&0&0&0&0
\\0&0&0&0&6&0&0&0
\\0&0&0&0&0&6&0&0
\\0&0&0&0&0&0&10&0
\\0&0&0&0&0&0&0&10\end{array}\right).
\end{equation}
On the diagonal the numbers $\frac{1}{2}(l+1)(l+2)$ appear for $l=0,1,2,...$. The crucial identities (3.11) and (3.16) are proved by use of the generating function of the Legendre polynomials, or by use of known identities relating the polynomials.

\section{\label{IV}Linear chain problem}

The question arises how to maximize the mean swimming velocity for given mean rate of
dissipation. This leads to an eigenvalue problem for the set of multipole coefficients $\vc{\mu}$,
\begin{equation}
\label{4.1}\du{B}\vc{\mu}_\lambda=\lambda\du{A}\vc{\mu}_\lambda.
\end{equation}
The mathematical discussion is simplified by truncating the matrices at a maximum $l$-value,
say $L$. We call the truncated $2L+2$-dimensional matrices $\du{A}_{0L}$ and $\du{B}_{0L}$.
The truncated matrices correspond to swimmers obeying the constraint that all multipole coefficients for $l>L$ vanish.

It is seen from Eq. (3.12) that there is a degeneracy in the
problem. The sum for the mean velocity consists of a sum of two
interlaced chains. In the one chain the $s$-coefficients for even
$l$ and the $c$-coefficients for odd $l$ appear. In the other
chain the $s$-coefficients for odd $l$ and the $c$-coefficients
for even $l$ appear. It is therefore sufficient to consider the
first type of chain. Eigenvectors of this form with the
coefficients for the second chain put equal to zero can be mapped
onto eigenvectors for the same eigenvalue with the two chains
interchanged. We call eigenvectors of the first type even, and
eigenvectors of the second type odd. The degeneracy corresponds to
invariance under a shift in time by $\pi/2\omega$.

There is also a symmetry under time reversal. Eigenvalues appear in pairs $\pm\lambda_j$.
The even eigenvector for $-\lambda_j$ can be obtained from the even eigenvector for $+\lambda_j$
by the replacement of the $c$-coefficients by their opposites, leaving the $s$-coefficients unchanged.
For the two conjugate eigenvectors the swimming velocity is equal and opposite for the same rate of dissipation.

The first symmetry allows a simplification of the eigenvalue problem by a reduction of the matrix dimension by a factor one half. There is a duplication in the matrices $\du{B}$ and $\du{A}$ which can be removed by use of complex notation. Thus we introduce the complex multipole moment
\begin{equation}
\label{4.2}\mu^c_l=(-i)^l(\mu_{lc}+i\mu_{ls}),
\end{equation}
and correspondingly instead of Eq. (3.13)
\begin{equation}
\label{4.3}\vc{\mu}^c=(\mu^c_{0},\mu^c_{1},\mu^c_{2},....).
\end{equation}
Then $\overline{U}_2$ and $\overline{\mathcal{D}}_2$ can be expressed as
\begin{equation}
\label{4.4}\overline{U_2}=\frac{1}{2}\omega
a(\vc{\mu}^c|\du{B}^c|\vc{\mu}^c),\qquad\overline{\mathcal{D}_2}=8\pi\eta\omega^2a^3
(\vc{\mu}^c|\du{A}^c|\vc{\mu}^c),
\end{equation}
with the notation
\begin{equation}
\label{4.5}(\vc{\mu}^c|\du{B}^c|\vc{\mu}^c)=\sum^\infty_{ll'}\mu^{c*}_lB^c_{ll'}\mu^c_{l'}.
\end{equation}
The truncated matrices $\du{B}^c_{03}$ and $\du{A}^c_{03}$ read
\begin{equation}
\label{4.6}\du{B}^c_{03}=\left(\begin{array}{cccc}
0&1&0&0
\\1&0&3&0
\\0&3&0&6
\\0&0&6&0
\end{array}\right),\qquad\du{A}^c_{03}=\left(\begin{array}{cccc}
1&0&0&0
\\0&3&0&0
\\0&0&6&0
\\0&0&0&10
\end{array}\right).
\end{equation}
The eigenvalue problem now reads
\begin{equation}
\label{4.7}\du{B}^c|\vc{\mu}^c_\lambda)=\lambda\du{A}^c|\vc{\mu}^c_\lambda).
\end{equation}
Since the matrices $\du{B}^c$ and $\du{A}^c$ are real and symmetric, the eigenvectors can be chosen to be real.

With truncation at $l=L$ the eigenvalue problem Eq. (4.7) is identical to that for a linear harmonic chain with masses corresponding to the diagonal elements of the matrix $\du{A}^c_{0L}$ and spring constants corresponding to the off-diagonal elements of the matrix $\du{B}^c_{0L}$. We can simplify further by renormalizing such that the masses are equal. Thus we introduce the modified moments
\begin{equation}
\label{4.8}f_l=\sqrt{(l+1)(l+2)}\mu^c_l.
\end{equation}
With these moments the rate of dissipation is
\begin{equation}
\label{4.9}\overline{\mathcal{D}_2}=4\pi\eta\omega^2a^3\sum^L_{l=0}|f_l|^2
=8\pi\eta\omega^2a^3(\vc{f}|\du{A}^{c\prime}|\vc{f}),
\end{equation}
where $\du{A}^{c\prime}=\frac{1}{2}\du{I}$ with unit matrix
$\du{I}$, and the swimming velocity is
\begin{equation}
\label{4.10}\overline{U_2}=\frac{1}{2}\omega
a\sum^L_{l=0}k_l\mathrm{Re}f_l^*f_{l+1}=\frac{1}{2}\omega
a(\vc{f}|\du{B}^{c\prime}|\vc{f}),
\end{equation}
where $\du{B}^{c\prime}$ is symmetric with non-zero elements
\begin{equation}
\label{4.11}B^{c\prime}_{l,l+1}=B^{c\prime}_{l+1,l}=\frac{1}{2}k_l,\qquad
k_l=\sqrt{\frac{l+1}{l+3}}.
\end{equation}
The coefficients $k_l$ tend to unity for large $l$, so that the eigenvalue problem
\begin{equation}
\label{4.12}
\du{B}^{c\prime}|\vc{f}_\lambda)=\lambda\du{A}^{c\prime}|\vc{f}_\lambda),
\end{equation}
corresponds to a chain of equal masses coupled by spring constants
which become uniform for large $l$.

We impose the constraint that the multipole coefficients for $l=0$
vanish. The coefficients for $l=0$ correspond to uniform spherical
expansion, which is excluded if we impose volume conservation. We
denote the matrices truncated at $L$ and with the first two rows
and columns deleted as $\du{A}_{1L}$ and $\du{B}_{1L}$. These have
dimension $2L$. The corresponding matrices $\du{A}^c_{1L}$ and
$\du{B}^c_{1L}$ have dimension $L$ and the matrices
$\du{A}^{c\prime}_{1L}$ and $\du{B}^{c\prime}_{1L}$ have dimension
$L$.

The eigenvalue problem Eq. (4.12) for the linear chain of $L$ equal masses coupled with equal force constants has eigenvalues
\begin{equation}
\label{4.13}
\lambda_q=2\cos\bigg(\frac{q\pi}{L+1}\bigg),\qquad q=1,...,L,
\end{equation}
and corresponding eigenvectors with components
\begin{equation}
\label{4.14}
f_{k,q}=C_q\sin\bigg(\frac{kq\pi}{L+1}\bigg),\qquad k,q=1,...,L,
\end{equation}
where $C_q$ is a normalization factor. The largest eigenvalue occurs for $q=1$. For this eigenvalue the components of the eigenvector vary slowly with $k$. In the limit $L\rightarrow\infty$ the maximum eigenvalue tends to $2$ and the components of the corresponding eigenvector tend to a constant.

\section{\label{V}Speed, power, efficiency}

As characteristic dimension of the sphere we take the diameter
$2a$. The dimensionless efficiency of translational swimming is
defined as the ratio \cite{7}
\begin{equation}
\label{5.1}E_T=4\eta\omega a^2\frac{|\overline{U_2}|}{\overline{\mathcal{D}_2}}.
\end{equation}
The optimum efficiency is related to the maximum eigenvalue by
\begin{equation}
\label{5.2}E_{T\mathrm{max}}=\lambda_{\mathrm{max}}/(4\pi).
\end{equation}
Due to a different normalization of the matrix $\du{B}$ the eigenvalue is four times that defined earlier \cite{8}.
It follows from Eq. (4.13) that the optimum efficiency is $1/(2\pi)$.
It is therefore of interest to consider the relative efficiency
\begin{equation}
\label{5.3}\eta_{1\mathrm{pot}}=2\pi E_T
\end{equation}
as a measure of efficiency in the space of potential flows. Here we have used the notation of Shapere and Wilczek \cite{5}.

We denote the eigenvector with largest eigenvalue of the truncated eigenvalue problem Eq. (4.7) with matrices $\du{A}^c_{1L}$ and $\du{B}^c_{1L}$ as $\vc{g}_{1L}$,
with normalization $(\vc{g}_{1L}|\vc{g}_{1L})=1$, and define
\begin{equation}
\label{5.4}\hat{U}_{1L}=(\vc{g}_{1L}|\du{B}^c_{1L}|\vc{g}_{1L}),\qquad\hat{\mathcal{D}}_{1L}
=(\vc{g}_{1L}|\du{A}^c_{1L}|\vc{g}_{1L}).
\end{equation}
Then correspondingly
\begin{equation}
\label{5.5}\frac{\hat{U}_{1L}}{\hat{\mathcal{D}}_{1L}}=\lambda_{\mathrm{max}}(1,L).
\end{equation}
The maximum eigenvalue $\lambda_{\mathrm{max}}(1,L)$
increases monotonically with $L$, since with increasing $L$  the space of possible modes gets larger. In Fig. 1 we plot $\frac{1}{2}\lambda_{\mathrm{max}}(1,L)$ for
values $L=2,...,30$. In Fig. 2 we show the components of the eigenvector $\vc{g}_{1L}$
with largest eigenvalue for $L=8$.

As shown in Fig. 1 the efficiency $E_{T\mathrm{max}}(1,L)$
increases monotonically with $L$. This suggests that the limit
$L\rightarrow\infty$ corresponds to the best swimmer. However, it
is worthwhile to consider also the dimensionless speed
$\hat{U}_{1L}$ and power $\hat{\mathcal{D}}_{1L}$ separately.
It is seen numerically that both quantities increase linearly with
$L$ at large $L$. When listing values for different $L$ we are
comparing speed and power for eigenvectors with the same
normalization.

It makes more sense to compare chains with the same amplitude of
motion. It follows from Eqs. (3.6) and (3.10) that for the
eigenvector $\vc{g}_{1L}$ the displacement $\vc{\xi}(t)$ at
$\theta=\pi/2$ describes an ellipse in the $zx$ plane given by the
equation
\begin{equation}
\label{5.6}\frac{\xi_x^2}{A(1,L)^2}+\frac{\xi_z^2}{B(1,L)^2}=a^2,
\end{equation}
with $A(1,L)$ and $B(1,L)$ given by
\begin{equation}
\label{5.7}A(1,L)=|\sum^L_{l=1}(l+1)\mathrm{Re}(i^lg_{1L,l})P_l(0)|,\qquad
B(1,L)=|\sum^L_{l=1}\mathrm{Im}(i^lg_{1L,l})P^1_l(0)|.
\end{equation}
For multipoles given by
$\vc{g}_{1L}/A(1,L)$ the ellipse described by $\vc{\xi}(t)$ will
have vertical semi-axis $a$ and horizontal semi-axis $b=B(1,L)a/A(1,L)$, if we take the $z$ axis to be horizontal. We find that the vertical semi-axis is larger than the horizontal one, except for $L=3$ and $L=5$.
For multipoles $\varepsilon\vc{g}_{1L}/A(1,L)$ the vertical semi-axis
has length $\varepsilon a$, where $\varepsilon$ can be taken to be
independent of $L$. We therefore consider the reduced speed and
power at fixed vertical amplitude of stroke,
\begin{equation}
\label{5.8}\hat{U}^A_{1L}=\frac{\hat{U}_{1L}}{A(1,L)^2},\qquad\hat{\mathcal{D}}^A_{1L}
=\frac{\hat{\mathcal{D}}_{1L}}{A(1,L)^2}.
\end{equation}
In Fig. 3 we plot the reduced speed $\hat{U}^A_{1L}$ as a function
of $L$, and in Fig. 4 we plot the reduced power
$\hat{\mathcal{D}}^A_{1L}$ as a function of $L$. Remarkably, the
reduced power at fixed amplitude shows a minimum at $L=8$, given
by $\hat{\mathcal{D}}^A_{18}=2.761$. An animalcule for which the
amplitude of motion is given by its structure, and for which the
relative amplitude of stroke is fixed, say at $\varepsilon=0.1$,
swims with least power for displacement $\vc{\xi}(\theta,t)$
determined by the set of multipoles
$\varepsilon\vc{g}_{1L}/A(1,L)$ with $L=8$. At $L=8$ the reduced
amplitude is $A(1,8)=2.572$, and the reduced speed is
$\hat{U}^A_{18}=4.380$.

For the set of multipoles $\vc{\mu}^c(1,L)=\varepsilon\vc{g}_{1L}/A(1,L)$ the mean speed and rate of dissipation are
\begin{equation}
\label{5.9}\overline{U}_2=\frac{1}{2}\omega a\varepsilon^2\hat{U}^A_{1L},\qquad\overline{\mathcal{D}}_2
=8\pi\eta\omega^2 a^3\varepsilon^2\hat{\mathcal{D}}^A_{1L}.
\end{equation}
In low Reynolds number swimming the speed is proportional to the
power. It is incorrect to estimate the required power on the basis
of Stokes' law \cite{9}, which corresponds to pulling of the sphere
through the fluid. In the case of pulling the power is
proportional to the square of the speed.

For a bacterium of radius $0.1\;\mu\mathrm{m}$ in water of shear
viscosity $\eta=0.001$ in SI units, the power for $L=8$ is
$P=\overline{\mathcal{D}_2}=6.94\times
10^{-23}\varepsilon^2\omega^2$ watt. The corresponding speed is
$U=2.19\times 10^{-7}\varepsilon^2\omega$ m/sec. The frequency is estimated \cite{14} as $10^4$ sec$^{-1}$. This is to be compared with the viscous time
scale $\tau_v=a^2\rho/\eta=10^{-8}$ sec. The power is calculated from Eq. (5.9) as
$P=6.94\times 10^{-15}\varepsilon^2$ watt and speed
$U=2.19\times 10^{-3}\;\varepsilon^2$ m/sec. The efficiency is $E_T=0.126$, compared with the maximum possible for potential flow $E_{T\mathrm{max}}=1/(2\pi)=0.159$.

The metabolic rate of birds has been
measured as 20.000 watt/m$^3$, of which one quarter is estimated
to be available for mechanical work \cite{15}. Accepting the same
rate for bacteria, we have $P=2.09\times 10^{-17}$ watt, and hence
find relative amplitude $\varepsilon=0.055$ and speed $U=6.6\times
10^{-6}$ m/sec. Therefore the bacterium moves several
diameters per second, in reasonable agreement with experimental
data \cite{3}. The specific energy consumption, defined as the power divided by the product of speed and weight \cite{15}, is about five orders of magnitude larger than that of a Boeing 747. We note that Dusenbery \cite{9} estimates the
available power as only 3 watt/m$^3$, instead of 5000 watt/m$^3$.
In our calculation this low power level would lead to a much too
small speed.

\section{\label{VI}Time-dependent swimming}

It is of interest to study some features of the swimming motion in
more detail. As we have shown above, the mean speed and mean power
to second order in the displacement $\vc{\xi}(t)$ are given by
bilinear expressions derived from the first order flow pattern.
For a chosen characteristic amplitude the latter can be optimized
to provide speed at minimum power. The set of multipoles
$\varepsilon\vc{g}_{1L}/A(1,L)$ with $L=8$ corresponding to the
eigenvector with maximum eigenvalue leads to optimal swimming. In
Fig. 5 we plot the nearly circular motion of the displacement
vector at $\theta=3\pi/12,\theta=5\pi/12,\,\pi/2,\;7\pi/12,\theta=9\pi/12,$ and $\varepsilon=0.1$
for seven-eighth of the period $T=2\pi/\omega$ , starting at
$t=0$. In Fig. 6 we show the radial displacement as a function of
the polar angle $\theta$ at times $t=0,\;t=T/8$ and $t=T/4$. This
demonstrates the running wave character of the surface wave. The
plot for the tangential displacement looks similar.

The second order velocity $U_2(t)$ follows from Eq. (2.11). This
can be evaluated by use of Eq. (3.11), which yields
\begin{eqnarray}
\label{6.1}U_2(t)=\omega
a\sum^\infty_{l=0}(l+1)(l+2)&\big[&\mu_{lc}\mu_{l+1,s}\cos^2\omega
t
-\mu_{ls}\mu_{l+1,c}\sin^2\omega t+\nonumber\\
+&(&\mu_{ls}\mu_{l+1,s}-\mu_{lc}\mu_{l+1,c})\sin\omega t\cos\omega
t\; \big].
\end{eqnarray}
The time-average of this expression equals that given in Eq.
(3.12). In Fig. 7 we plot the ratio $U_2(t)/\overline{U}_2$ for
the optimal stroke with displacement $\vc{\xi}(\theta,t)$
determined by the set of multipoles
$\varepsilon\vc{g}_{1L}/A(1,L)$ with $L=8$. The maximum deviation
from unity is about one percent.

The second order rate of dissipation $\mathcal{D}_2(t)$ follows
from Eq. (2.13). This can be evaluated by use of Eq. (3.16), which
yields
\begin{equation}
\label{6.2}\mathcal{D}_2(t)=8\pi\eta\omega^2a^3
\sum^\infty_{l=0}(l+1)(l+2)\big[\mu_{lc}^2\cos^2\omega t
+\mu_{ls}^2\sin^2\omega t +2\mu_{lc}\mu_{ls}\sin\omega t\cos\omega
t\; \big].
\end{equation}
The time-average of this expression equals that given in Eq.
(3.17). In Fig. 7 we plot also the ratio
$\mathcal{D}_2(t)/\overline{\mathcal{D}}_2$ for the optimal stroke
with displacement $\vc{\xi}(\theta,t)$ determined by the set of
multipoles $\varepsilon\vc{g}_{1L}/A(1,L)$ with $L=8$. It turns out that for this stroke $\mathcal{D}_2(t)/\overline{\mathcal{D}}_2$
equals unity within numerical accuracy.

The second order flow velocity $\vc{v}_2(\vc{r},t)$ follows from
the second order velocity at the surface $\vc{u}_{2S}(\theta,t)$,
as given by Eq. (2.10). The latter can be expanded in terms of a
complete set of outgoing waves $\{\vc{v}^-_{l0\sigma}(\vc{r})\}$,
where $\sigma$ takes the values $0,2$, as indicated
elsewhere \cite{16}. The modes with $\sigma=0$ are accompanied by a
pressure disturbance. The contribution for $l=1,\;\sigma=0$ decays
with a long range flow pattern falling off as $1/r$. This must be
cancelled by a Stokes solution $\vc{v}_2^{St}(\vc{r},t)$ which
vanishes on the sphere of radius $a$ and tends to
$-U_2(t)\vc{e}_z$ as $r\rightarrow\infty$. The procedure can be
performed straightforwardly, but we shall not present the details.
In principle the perturbation expansion in powers of the surface
displacement, as indicated in Eq. (2.8), can be extended to higher
order in similar fashion.

\section{\label{VII}Axisymmetric polar flows}

In the following we extend the analysis to more general flows. We consider motions for which to first order in the
displacement the flow is axisymmetric and polar, so that in spherical coordinates $(r,\theta,\varphi)$ the flow velocity $\vc{v}(\vc{r},t)$ and the pressure $p(\vc{r},t)$
do not depend on $\varphi$, and $\vc{v}$ has vanishing component $v_\varphi$. In general the solutions of the Stokes equations for the flow about a sphere have been classified \cite{16} into three types indexed $\sigma=0,1,2$. The potential flows considered earlier are of type $\sigma=2$. We now consider in addition flows of type $\sigma=0$. For the potential flows the pressure disturbance vanishes, but the flows of type $\sigma=0$ cannot be expressed as the gradient of a scalar potential and there is a pressure disturbance. For an axisymmetric flow of type $\sigma=1$ the flow velocity has only a $v_\varphi$ component, and the pressure disturbance vanishes. Flows of this type do not contribute to the translational velocity of the sphere.

The first order flow outside the sphere is expanded as
\begin{equation}
\label{7.1}\vc{v}_1(r,\theta,t)=-\omega a\sum^\infty_{l=1}\bigg[\mu_l(t)\vc{u}_l(r,\theta)+\kappa_l(t)\vc{v}_l(r,\theta)\bigg],\qquad r>a,
\end{equation}
with component field $\vc{u}_l(r,\theta)$ given by Eq. (3.6), and $\vc{v}_l(r,\theta)$ given by
\begin{equation}
\label{7.2}
\vc{v}_l(r,\theta)=\bigg(\frac{a}{r}\bigg)^{l}\big[(l+1)P_l(\cos\theta)\vc{e}_r
+\frac{l-2}{l}P^1_l(\cos\theta)\vc{e}_\theta\big].
\end{equation}
In the second sum in Eq. (7.1) we must put $\kappa_1(t)=0$, since the term with $l=1$ would correspond to a force $F_1(t)\vc{e}_z$. We have normalized such that at $r=a$ the function $\vc{v}_l$ has the same radial component as $\vc{u}_l$.
The solution $\vc{u}_l(r,\theta)$ is of type $\sigma=2$, the solution $\vc{v}_l(r,\theta)$ is of type $\sigma=0$. The corresponding first order pressure is
\begin{equation}
\label{7.3}p_1(r,\theta,t)=-\omega a\sum^\infty_{l=2}\kappa_l(t)p_l(r,\theta),\qquad r>a
\end{equation}
with component pressure disturbance
\begin{equation}
\label{7.4}p_l(r,\theta)=2\eta (2l-1)a^l\Phi^-_l(r,\theta).
\end{equation}

The multipole coefficients $\mu_l(t)$ and $\kappa_l(t)$ in Eq.
(7.1) can be expressed as
\begin{equation}
\label{7.5}\mu_l(t)=\mu_{lc}\cos\omega t+\mu_{ls}\sin\omega t,\qquad\kappa_l(t)=\kappa_{lc}\cos\omega t+\kappa_{ls}\sin\omega t.
\end{equation}
The corresponding displacement is
\begin{eqnarray}
\label{7.6}\vc{\xi}(\theta,t)=a\sum^\infty_{l=1}&\big[&\big(\mu_{ls}\cos\omega
t-\mu_{lc}\sin\omega t\big)\vc{u}_l(a,\theta)\nonumber\\
&+&\big(\kappa_{ls}\cos\omega
t-\kappa_{lc}\sin\omega t\big)\vc{v}_l(a,\theta)\big].
\end{eqnarray}

In the calculation of the mean swimming velocity, as given by Eq.
(2.12), we use the identities
\begin{eqnarray}
\label{7.7}\int_{r=a}\vc{u}_k\cdot(\nabla\vc{u}_l)\cdot\vc{e}_z\;dS=&-&4\pi
k(k+1)a\delta_{k,l+1},\nonumber\\
\int_{r=a}\vc{u}_k\cdot(\nabla\vc{v}_l)\cdot\vc{e}_z\;dS=&-&8\pi
\frac{(k+1)(k+2)}{2k+3}\;a\delta_{k,l-1}\nonumber\\&-&4\pi
\frac{k(k+1)(2k-3)}{2k+1}\;a\delta_{k,l+1},\nonumber\\
\int_{r=a}\vc{v}_k\cdot(\nabla\vc{u}_l)\cdot\vc{e}_z\;dS=&-&4\pi
\frac{k(k+1(2k-1)}{2k+1})\;a\delta_{k,l+1},\nonumber\\
\int_{r=a}\vc{v}_k\cdot(\nabla\vc{v}_l)\cdot\vc{e}_z\;dS=&-&8\pi
\frac{(k+1)(k+2)(2k-1)}{(2k+1)(2k+3)}\;a\delta_{k,l-1}\nonumber\\&-&4\pi
\frac{k(k+1)(2k-3)^2}{(2k-1)(2k+1)}\;a\delta_{k,l+1}.
\end{eqnarray}
The first one is equivalent to Eq. (3.11). It follows that the mean swimming velocity is again given by a sum of products of adjacent multipole coefficients,
\begin{eqnarray}
\label{7.8}\overline{U_2}&=&\frac{1}{2}\omega
a\sum^\infty_{l=1}\bigg[(l+1)(l+2)\big[\mu_{lc}\mu_{l+1,s}-\mu_{ls}\mu_{l+1,c}\big]\nonumber\\
&+&\frac{(l+1)(l+2)(2l-1)}{2l+3}\big[\kappa_{lc}\mu_{l+1,s}-\kappa_{ls}\mu_{l+1,c}\big]\nonumber\\
&+&\frac{(l+1)(l+2)(2l-1)}{2l+3}\big[\mu_{lc}\kappa_{l+1,s}-\mu_{ls}\kappa_{l+1,c}\big]\nonumber\\
&+&(l+1)(l+2)\frac{(2l-3)(2l-1)}{(2l+1)(2l+3)}\big[\kappa_{lc}\kappa_{l+1,s}-\kappa_{ls}\kappa_{l+1,c}\big]\bigg].
\end{eqnarray}
We define the complex multipole moment vector $\vc{\psi}$ as the one-dimensional array
\begin{equation}
\label{7.9}\vc{\psi}=(\kappa_{1c}+i\kappa_{1s},\mu_{1c}+i\mu_{1s},\kappa_{2c}+i\kappa_{2s},\mu_{2c}+i\mu_{2s},....).
\end{equation}
Then $\overline{U_2}$ can be expressed as
\begin{equation}
\label{7.10}\overline{U_2}=\frac{1}{2}\omega
a(\vc{\psi}|\du{B}|\vc{\psi}),
\end{equation}
with a dimensionless pure imaginary and antisymmetric matrix
$\du{B}$. The upper left-hand corner of the matrix $\du{B}$,
truncated at $l=4$, reads
\begin{equation}
\label{7.11}\du{B}_{14}=i\left(\begin{array}{cccccccc}0&0&\frac{1}{5}&-\frac{3}{5}&0&0&0&0
\\0&0&-\frac{3}{5}&-3&0&0&0&0
\\-\frac{1}{5}&\frac{3}{5}&0&0&-\frac{18}{35}&-\frac{18}{7}&0&0
\\\frac{3}{5}&3&0&0&-\frac{18}{7}&-6&0&0
\\0&0&\frac{18}{35}&\frac{18}{7}&0&0&-\frac{50}{21}&-\frac{50}{9}
\\0&0&\frac{18}{7}&6&0&0&-\frac{50}{9}&-10
\\0&0&0&0&\frac{50}{21}&\frac{50}{9}&0&0
\\0&0&0&0&\frac{50}{9}&10&0&0\end{array}\right).
\end{equation}
We can impose the constraint $\kappa_1=0$ by dropping the first
element of $\vc{\psi}$ and erasing the first row and column of the
matrix $\du{B}$. We denote the corresponding modified vector as
$\hat{\vc{\psi}}$ and the modified matrix as $\hat{\du{B}}$.

The rate of dissipation $\mathcal{D}_2(t)$ is
expressed as a surface integral \cite{7}
\begin{equation}
\label{7.12}\mathcal{D}_2=-\int_{r=a}\vc{v}_{1}.\vc{\sigma}_{1}.\vc{e}_r\;dS,
\end{equation}
where $\vc{\sigma}_1$ is the first order stress tensor, given by
\begin{equation}
\label{7.13}\vc{\sigma}_{1}=\eta(\nabla\vc{v}_1+\widetilde{\nabla\vc{v}_1})-p_1\vc{I}.
\end{equation}
In the calculation of the rate of dissipation we use the identities
\begin{eqnarray}
\label{7.14}\int_{r=a}\vc{u}_k\cdot(\nabla\vc{u}_l)\cdot\vc{e}_r\;dS&=&-4\pi a
(k+1)(k+2)\delta_{kl},\nonumber\\
\int_{r=a}\vc{u}_k\cdot(\nabla\vc{v}_l+\widetilde\nabla\vc{v}_l-p_l)\cdot\vc{e}_r\;dS&=&-8\pi a
\frac{(k+1)(k+2)(2k-1)}{2k+1}\delta_{kl},\nonumber\\
\int_{r=a}\vc{v}_k\cdot(\nabla\vc{u}_l)\cdot\vc{e}_r\;dS&=&-4\pi a
\frac{(k+1)(k+2)(2k-1)}{2k+1}\delta_{kl},\nonumber\\
\int_{r=a}\vc{v}_k\cdot(\nabla\vc{v}_l+\widetilde\nabla\vc{v}_l-p_l)\cdot\vc{e}_r\;dS&=&-8\pi a
\frac{(k+1)(2k^3+k^2-2k+2)}{k(2k+1)}\delta_{kl}.\nonumber\\
\end{eqnarray}
The first one is equivalent to Eq. (3.16).
The time-averaged rate of dissipation is given by
\begin{eqnarray}
\label{7.15}\overline{\mathcal{D}_2}=8\pi\eta\omega^2a^3\sum^\infty_{l=1}&\bigg[&\frac{1}{2}(l+1)(l+2)(\mu_{lc}^2+\mu_{ls}^2)\nonumber\\
&+&\frac{(l+1)(l+2)(2l-1)}{2l+1}(\mu_{lc}\kappa_{lc}+\mu_{ls}\kappa_{ls})\nonumber\\
&+&\frac{(l+1)(2l^3+l^2-2l+2)}{2l(2l+1)}(\kappa_{lc}^2+\kappa_{ls}^2)\bigg].
\end{eqnarray}
This can be expressed as
\begin{equation}
\label{7.16}\overline{\mathcal{D}_2}=8\pi\eta\omega^2a^3(\vc{\psi}|\du{A}|\vc{\psi}),
\end{equation}
with a dimensionless real and symmetric matrix $\du{A}$. We denote
the modified matrix obtained by dropping the first row and column
by $\hat{\du{A}}$. The upper left-hand corner of the matrix
$\du{A}$, truncated at $l=4$, reads
\begin{equation}
\label{7.17}\du{A}_{14}=\left(\begin{array}{cccccccc}1&1&0&0&0&0&0&0
\\1&3&0&0&0&0&0&0
\\0&0&\frac{27}{10}&\frac{18}{5}&0&0&0&0
\\0&0&\frac{18}{5}&6&0&0&0&0
\\0&0&0&0&\frac{118}{21}&\frac{50}{7}&0&0
\\0&0&0&0&\frac{50}{7}&10&0&0
\\0&0&0&0&0&0&\frac{115}{12}&\frac{35}{3}
\\0&0&0&0&0&0&\frac{35}{3}&15\end{array}\right).
\end{equation}
If the elements corresponding to the multipole moments $\{\kappa_l\}$ are omitted, then these results reduce to those obtained earlier for irrotational flows.

\section{\label{VIII}Optimization for axisymmetric polar flows}

We impose the constraint that the force exerted on the fluid vanishes at any time. This requires $\kappa_1(t)=0$. With
this constraint the mean swimming velocity
$\overline{U}_2$ and the mean rate of dissipation $\overline{\mathcal{D}}_2$ can be expressed as
\begin{equation}
\label{8.1}\overline{U_2}=\frac{1}{2}\omega
a(\hat{\vc{\psi}}|\hat{\du{B}}|\hat{\vc{\psi}}),\qquad\overline{\mathcal{D}_2}=8\pi\eta\omega^2a^3
(\hat{\vc{\psi}}|\hat{\du{A}}|\hat{\vc{\psi}}).
\end{equation}
Optimization of the mean swimming velocity for given mean rate of dissipation leads to the eigenvalue problem
\begin{equation}
\label{8.2}\hat{\du{B}}|\hat{\vc{\psi}}_\lambda)=\lambda\hat{\du{A}}|\hat{\vc{\psi}}_\lambda).
\end{equation}
The matrix $\hat{\du{B}}$ is pure imaginary and antisymmetric and
the matrix $\hat{\du{A}}$ is real and symmetric. As in the case of
potential flows we truncate at maximum $l$-value $L$. The
 truncated matrices $\hat{\du{A}}_{1L}$ and $\hat{\du{B}}_{1L}$ are $2L-1$-dimensional. The
 structure of the eigenvalue equations is such that they can be satisfied for
 real eigenvalues by eigenvectors with components which are real for odd $l$
 and pure imaginary for even $l$. The complex conjugate of an eigenvector
 corresponds to the eigenvalue for the opposite sign. Hence it suffices to consider the positive eigenvalues.
 In our plots we have chosen the phase of the eigenvectors such that the first potential multipole moment $\mu^c_1$ is real and positive.

With truncation at $l=L$ the eigenvalue problem is equivalent to
that for two coupled linear harmonic chains with masses
corresponding to the diagonalized form of the matrix
$\hat{\du{A}}_{1L}$. However, it is not necessary to perform this
diagonalization explicitly, and it suffices to discuss Eq. (8.2)
directly. It is of interest to consider the $2\times 2$ matrix
along the diagonal direction of the matrix $\du{A}$ for large $l$.
Diagonalization of this matrix shows that one of its eigenvalues
is of order unity, whereas the second one grows as $l^2$ as $l$
increases. For the eigenvector corresponding to the first
eigenvalue the second component is nearly the opposite of the
first, and for the second eigenvalue the two components are nearly
equal. This suggests that the eigenvector with largest eigenvalue
for the problem Eq. (8.2) for large $L$ is a mixture of flows of
potential and viscous type with nearly equal and opposite
amplitudes. This is confirmed by numerical solution of the
eigenvalue problem for a large value of $L$, say $L=40$. If the
optimal eigenvector is decomposed into potential and viscous
components, corresponding to $\mu$- and $\kappa$-moments
respectively,
\begin{equation}
\label{8.3}\hat{\vc{\psi}}_\lambda=\hat{\vc{\psi}}_{\lambda
p}+\hat{\vc{\psi}}_{\lambda v},
\end{equation}
then the norm of the viscous part is nearly equal to the norm of
the potential part.

It turns out that the inclusion of the viscous part has a dramatic
effect on the maximum eigenvalue. In Fig. 8 we show the maximum
eigenvalue as a function of $L$, in analogy with Fig. 1. This
shows that $\lambda_{max}$ tends to a constant larger than 2 for
large $L$.

We prove that the constant equals $2\sqrt{2}$. The inclusion of viscous flows has led to a qualitative change. It is
no longer sufficient to consider the asymptotically uniform linear chain as in Sec. IV. The asymptotic variation of couplings
and masses along two coupled linear harmonic chains must be taken into account. With modified moments as in Eq. (4.8)
\begin{equation}
\label{8.4}f_l=\sqrt{(l+1)(l+2)}\mu_l,\qquad
g_l=\sqrt{(l+1)(l+2)}\kappa_l,
\end{equation}
the $6\times 6$ matrices along the diagonal of the corresponding
matrices $\du{A}'$ and $\du{B}'$ linking the multipoles of order
$l-1,l$ and $l+1$ in the limit of large $l$ take the form
\begin{equation}
\label{8.5}\du{A}'^{(6)}_0=\frac{1}{2}\left(\begin{array}{cccccc}
1&1&0&0&0&0
\\1&1&0&0&0&0
\\0&0&1&1&0&0
\\0&0&1&1&0&0
\\0&0&0&0&1&1
\\0&0&0&0&1&1
\end{array}\right),\qquad\du{B}'^{(6)}_0=\frac{1}{2}\left(\begin{array}{cccccc}
0&0&-i&-i&0&0
\\0&0&-i&-i&0&0
\\i&i&0&0&-i&-i
\\i&i&0&0&-i&-i
\\0&0&i&i&0&0
\\0&0&i&i&0&0
\end{array}\right).
\end{equation}
In comparison the large $l$ behavior of the $3\times 3$ matrices
along the diagonal of the matrices $\du{A}^{c\prime}$ and
$\du{B}^{c\prime}$ of Sec. 4 is given by
\begin{equation}
\label{8.6}\du{A}^{c\prime(3)}_0=\frac{1}{2}\left(\begin{array}{ccc}
1&0&0
\\0&1&0\\0&0&1
\end{array}\right),\qquad\du{B}^{c\prime(3)}_0=\frac{1}{2}\left(\begin{array}{ccc}
0&1&0\\1&0&1
\\0&1&0
\end{array}\right).
\end{equation}
The eigenvalue problem
$\du{B}^{c\prime(3)}_0|\vc{f}^{(3)})=\lambda\du{A}^{c\prime(3)}_0|\vc{f}^{(3)})$
has eigenvalues $\lambda_{0\pm}=\pm\sqrt{2}$, $\lambda_{00}=0$, and
the eigenvalue problem
$\du{B}'^{(6)}_0|\vc{f}^{(6)})=\lambda\du{A}'^{(6)}_0|\vc{f}^{(6)})$
has the same eigenvalues, each twofold degenerate. However, the result is unstable under small perturbations, and the higher order terms
of the matrix elements to order $1/l^2$ must be considered to
obtain the correct result corresponding to the coupled linear chains.

Thus instead of Eq. (8.5) we consider the asymptotic behavior
\begin{eqnarray}
\label{8.7}\du{A}'^{(6)}(l)&=&\du{A}'^{(6)}_0+\du{A}'^{(6)}_1\frac{1}{l}+\du{A}'^{(6)}_2\frac{1}{l^2}
+O\big(\frac{1}{l^3}\big),\nonumber\\
\du{B}'^{(6)}(l)&=&\du{B}'^{6)}_0+\du{B}'^{(6)}_1\frac{1}{l}+\du{B}'^{(6)}_2\frac{1}{l^2}
+O\big(\frac{1}{l^3}\big).
\end{eqnarray}
From Eqs. (7.8) and (7.15) one finds that the matrices
$\du{A}'^{(6)}_1$ and $\du{B}'^{(6)}_1$ are given by
\begin{eqnarray}
\label{8.8}\du{A}'^{(6)}_1&=&\frac{1}{2}\left(\begin{array}{cccccc}
-2&-1&0&0&0&0
\\-1&0&0&0&0&0
\\0&0&-2&-1&0&0
\\0&0&-1&0&0&0
\\0&0&0&0&-2&-1
\\0&0&0&0&-1&0
\end{array}\right),\nonumber\\
\du{B}'^{(6)}_1&=&\frac{-i}{2}\left(\begin{array}{cccccc}
0&0&-5&-3&0&0
\\0&0&-3&-1&0&0
\\5&3&0&0&-5&-3
\\3&1&0&0&-3&-1
\\0&0&5&3&0&0
\\0&0&3&1&0&0
\end{array}\right).
\end{eqnarray}
The matrices $\du{A}'^{(6)}_2$ and $\du{B}'^{(6)}_2$ are given by
\begin{eqnarray}
\label{8.9}\du{A}'^{(6)}_2&=&\frac{1}{4}\left(\begin{array}{cccccc}
2&-1&0&0&0&0
\\-1&0&0&0&0&0
\\0&0&6&1&0&0
\\0&0&1&0&0&0
\\0&0&0&0&10&3
\\0&0&0&0&3&0
\end{array}\right),\nonumber\\
\du{B}'^{(6)}_2&=&\frac{-i}{4}\left(\begin{array}{cccccc}
0&0&19&9&0&0
\\0&0&9&3&0&0
\\-19&-9&0&0&29&15
\\-9&-3&0&0&15&5
\\0&0&-29&-15&0&0
\\0&0&-15&-5&0&0
\end{array}\right).
\end{eqnarray}
From the eigenvalue equation
$|\du{B}'^{(6)}_0+z\du{B}'^{(6)}_1-\lambda_1(\du{A}'^{(6)}_0+z\du{A}'^{(6)}_1)|=0$ one finds that in the
limit $z\rightarrow 0$ the eigenvalues tend to $\lambda_{1\pm}=\pm 2\sqrt{2},\lambda_{10}=0$, each twofold degenerate.
From the eigenvalue equation
$|\du{B}'^{(6)}_0+z\du{B}'^{(6)}_1+z^2\du{B}'^{(6)}_2-\lambda_2(\du{A}'^{(6)}_0+z\du{A}'^{(6)}_1+z^2\du{A}'^{(6)}_2)|=0$ one finds that in the
limit $z\rightarrow 0$ the eigenvalues tend to $\lambda_{2\pm}=\pm 2,\lambda_{20}=0$, each twofold degenerate. The largest eigenvalue $\lambda_{2+}=2$ is a factor $\sqrt{2}$ larger than $\lambda_{0+}=\sqrt{2}$ given below Eq. (8.6). Hence for the complete problem with matrices $\du{B}_{1L}$ and $\du{A}_{1L}$ the maximum eigenvalue in the limit $L\rightarrow\infty$ is a factor $\sqrt{2}$ larger than obtained from the linear chain problem for potential flows of Sec. IV. The maximum eigenvalue for the present problem therefore tends to $2\sqrt{2}$ in the limit $L\rightarrow\infty$, as suggested by Fig. 8.

Thus with the inclusion of $\sigma=0$ modes the efficiency of translational swimming defined in Eq. (5.1) takes the maximum value
\begin{equation}
\label{8.10}E_{T\mathrm{max}}=\frac{1}{\pi\sqrt{2}}.
\end{equation}
As in the case of potential swimming the optimum value is reached for a set of multipoles decaying in absolute magnitude as $1/l$ at large $l$. This suggests that the maximization of $E_T$ leads to an optimum stroke which is not of physical relevance.

\section{\label{IX}Speed and power}

We denote the eigenvector with maximum eigenvalue corresponding to the truncated matrices $\hat{\du{A}}_{1L}$ and $\hat{\du{B}}_{1L}$ as $\vc{g}_{1L}$ with normalization $(\vc{g}_{1L}|\vc{g}_{1L})=1$. As in Sec. V we look for a different selection criterion for optimization of the stroke.

For the more general axisymmetric flow patterns we find again that for the eigenvector $\vc{g}_{1L}$ the displacement $\vc{\xi}(t)$ at $\theta=\pi/2$ describes an ellipse in the $zx$ plane given by Eq. (5.6), but now with modified expressions for the coefficients $A(1,L)$ and $B(1,L)$. More generally we consider arbitrary values of $\theta$. We then find that in general the vector $\vc{\xi}(t)$ describes an ellipse in the $zx$ plane which is tilted with respect to the $z$ axis. The shape and tilt of the ellipse are described conveniently by Stokes parameters \cite{17}.

The components $\xi_z(\theta,t)$ and $\xi_x(\theta,t)$ can be expressed as
\begin{equation}
\label{9.1}\xi_z(\theta,t)=\mathrm{Im}\big(\alpha_L(\theta) e^{-i\omega t}\big)a,\qquad\xi_x(\theta,t)=\mathrm{Im}\big(\beta_L(\theta) e^{-i\omega t}\big)a,
\end{equation}
with complex amplitudes $\alpha_L(\theta)$ and $\beta_L(\theta)$ given by
\begin{eqnarray}
\label{9.2}\alpha_L(\theta)&=&q_L(\theta)\cos\theta-p_L(\theta)\sin\theta,\nonumber\\
\beta_L(\theta)&=&q_L(\theta)\sin\theta+p_L(\theta)\cos\theta,
\end{eqnarray}
where
 \begin{eqnarray}
\label{9.2}p_L(\theta)&=&\sum^L_{l=1}g_{1L,2l-1}P^1_l(\cos\theta)+\sum^{L-1}_{l=1}g_{1L,2l}\frac{l-1}{l+1}P^1_{l+1}(\cos\theta),\nonumber\\
q_L(\theta)&=&\sum^L_{l=1}g_{1L,2l-1}(l+1)P_l(\cos\theta)+\sum^{L-1}_{l=1}g_{1L,2l}(l+2)P_{l+1}(\cos\theta).
\end{eqnarray}
The Stokes parameters of the ellipse at polar angle $\theta$ are defined by \cite{17}
 \begin{eqnarray}
\label{9.3}I_S&=&|\alpha|^2+|\beta|^2,\qquad Q_S=|\alpha|^2-|\beta|^2,\qquad \delta=\mathrm{arg}\frac{\alpha}{\beta},\nonumber\\
U_S&=&2|\alpha||\beta|\cos\delta,\qquad V_S=2|\alpha||\beta|\sin\delta,
\end{eqnarray}
where for brevity we have omitted the subscript $L$ and the variable $\theta$.
The tilt angle of the ellipse is given by
\begin{equation}
\label{9.4}\gamma_S=\frac{1}{2}\arctan\frac{U_S}{Q_S},
\end{equation}
and the ellipticity $\varepsilon_S$ follows from
\begin{equation}
\label{9.5}\eta_S=\frac{1}{2}\arctan\frac{V_S}{\sqrt{Q_S^2+V_S^2}},\qquad\varepsilon_S=|\tan\eta_S|.
\end{equation}
The long and short semi-axis of the ellipse are
\begin{equation}
\label{9.6}P(1,L,\theta)=\sqrt{\frac{I_S}{1+\varepsilon_S^2}},\qquad Q(1,L,\theta)=\varepsilon_SP(1,L,\theta).
\end{equation}

We find for each $L$ that the ellipse described by $\vc{\xi}(t)$ at $\theta=\pi/2$ for the stroke with maximum efficiency $E_T$ has its long axis parallel to the $z$ axis. Thus if we represent the ellipse again by Eq. (5.6) then for multipoles given by $\vc{g}_{1L}/B(1,L)$ the ellipse described by $\vc{\xi}(t)$ will have horizontal semi-axis $a$ and vertical semi-axis $b=A(1,L)a/B(1,L)$. We therefore consider the reduced speed and power at fixed horizontal amplitude of stroke,
  \begin{equation}
\label{9.7}\hat{U}^B_{1L}=\frac{\hat{U}_{1L}}{B(1,L)^2},\qquad\hat{\mathcal{D}}^B_{1L}
=\frac{\hat{\mathcal{D}}_{1L}}{B(1,L)^2},
\end{equation}
with
  \begin{equation}
\label{9.8}B(1,L)=P(1,L,\frac{\pi}{2}).
\end{equation}
In Fig. 9 we
show the plot of $\hat{U}^B_{1L}$ for the optimal eigenvector as a
function of $L$, and in Fig. 10 we show the corresponding plot for
the reduced power $\hat{D}^B_{1L}$. The reduced power shows again
a minimum, this time at $L=7$, given by $\hat{D}^B_{17}=1.529$. At $L=7$ the reduced amplitude is $B(1,7)=2.945$, and the reduced speed is $\hat{U}^B_{17}=3.303$.  In Fig. 11 we plot the absolute values of the set of multipole moments $\{\kappa_l,\mu_l\}$ for the optimal eigenvector with $L=7$.

For the set of complex multipoles $\hat{\vc{\psi}}(1,L)=\varepsilon\vc{g}_{1L}/B(1,L)$ the mean speed and rate of dissipation are
\begin{equation}
\label{9.9}\overline{U}_2=\frac{1}{2}\omega a\varepsilon^2\hat{U}^B_{1L},\qquad\overline{\mathcal{D}}_2
=8\pi\eta\omega^2 a^3\varepsilon^2\hat{\mathcal{D}}^B_{1L}.
\end{equation}
Performing the same estimate as at the end of Sec. V for the more general class of flows with the optimum stroke for $L=7$ we find power $P=3.84\times 10^{-15}\varepsilon^2$ watt and speed $U=1.65\times 10^{-3}\varepsilon^2$ m/sec. The efficiency is $E_T=0.172$, compared with the maximum possible for general flow $E_{T\mathrm{max}}=1/(\pi\sqrt{2})=0.225$. For $P=2.09\times 10^{-17}$ watt we find relative amplitude $\varepsilon=0.074$ and speed $U=9.0\times 10^{-6}$ m/sec.

The nature of the optimum stroke for $L=7$ is shown in Fig. 12, in analogy to Fig. 5. The time-dependent swimming velocity $U_2(t)$ and rate of dissipation $D_2(t)$ can be evaluated in analogy to Eqs. (6.1) and (6.2). The dimensionless ratios $U_2(t)/\overline{U}_2$ and $D_2(t)/\overline{D}_2$ for the optimal stroke with $L=7$ vary in time quite similarly to the behavior shown in Fig. 7. Again the ratio $D_2(t)/\overline{D}_2$ equals unity within numerical accuracy.

\section{\label{X}Discussion}

Basing ourselves on the Stokes equations, rather than the linearized Navier-Stokes equations, we have developed a simpler discussion of the swimming of a sphere at low Reynolds number with the restriction to potential flow solutions than was presented before \cite{8}. The identities Eqs. (3.11) and (3.16) play a crucial role. They imply that the representation of the flow in terms of electrostatic multipole potentials is particularly simple. In this representation the matrix $\du{A}^c$, from which the rate of dissipation is calculated, is diagonal, and the matrix $\du{B}^c$, from which the swimming velocity is calculated, is tri-diagonal. Correspondingly, the eigenvalue problem which yields the swimming stroke of maximum efficiency, is relatively simple.

Subsequently we have extended the derivation to the complete set of axisymmetric polar solutions of the Stokes equations. An additional set of multipole moments corresponding to flows with vorticity needs to be introduced. Although this leads to a doubling of dimensionality, the structure of the eigenvalue problem in the chosen representation remains fairly simple.

The additional flow solutions allow a considerable enhancement of efficiency, defined as the dimensionless ratio of speed and power. As in the case of irrotational flow, the maximum efficiency is attained for a stroke characterized by multipoles with a significant weight at high order. This indicates that the efficiency is not the most suitable measure of swimming performance.

Therefore we have considered a measure of performance based on a comparison of energy consumption for strokes with the same amplitude. The measure allows selection of a stroke with minimum energy consumption in a class of possible strokes. The optimal stroke selected in this manner involves multipoles of relatively low order and is expected to be of physical interest.

Although the spherical geometry provides only a crude approximation to the shape of most microorganisms, it has the advantage that the mechanism of swimming can be analyzed in great detail. The analysis shows that it is worthwhile to consider various measures of swimming performance. The mathematical formalism may serve as a guide in the study of more complicated geometry, such as a spheroid or an ellipsoid.

\newpage

\section*{Figure captions}

\subsection*{Fig. 1}
Plot of one-half the maximum eigenvalue
$\frac{1}{2}\lambda_\mathrm{max}(1,L)$ for sets of multipoles $\{\mu_{ls},\mu_{lc}\}$
with $1\leq l\leq L$ as a function of $L$ for $L=2,...,30$.  The values tend to unity as $L\rightarrow\infty$.

\subsection*{Fig. 2}
Plot of the components of the eigenvector with
largest eigenvalue, normalized to unity, for $L=8$. The corresponding multipoles $\{\mu_{ls},\mu_{lc}\}$
with $1\leq l\leq 8$ follow from Eq. (4.2). The values $\mu_{ls}$ for $l$ even and the values $\mu_{lc}$ for $l$ odd vanish.

\subsection*{Fig. 3}
Plot of the reduced speed $\hat{U}^A_{1L}$ for fixed maximum
amplitude of the displacement at $\theta=\pi/2$ as a function of
$L$. At each value of $L$ the most efficient set of multipoles $\{\mu_{1s},\mu_{1c},...,\mu_{Ls},\mu_{Lc}\}$ for swimming via irrotational flow is considered.

\subsection*{Fig. 4}
Plot of the reduced power $\hat{\mathcal{D}}^A_{1L}$ for fixed
maximum amplitude of the displacement at $\theta=\pi/2$ as a
function of $L$. At each value of $L$ the most efficient set of multipoles $\{\mu_{1s},\mu_{1c},...,\mu_{Ls},\mu_{Lc}\}$ for swimming via irrotational flow is considered.

\subsection*{Fig. 5}
Plot of the end of the displacement vector $\vc{\xi}(t)$ at
$\theta=3\pi/12,5\pi/12,\;\pi/2,7\pi/12$ and $9\pi/12$ for maximum amplitude of
the displacement at $\theta=\pi/2$ equal to $0.1\;a$ for the optimum eigenvector for $L=8$ with complex multipoles $\{\mu^c_l\}$. The motion
is depicted with start at $t=0$ and finish at $t=\frac{7}{8}T$, where
$T=2\pi/\omega$. The endpoint is marked by a small circle.

\subsection*{Fig. 6}
Plot of the radial displacement for maximum amplitude $0.1\;a$ as
a function of polar angle $\theta$ for $t=0$ (solid curve), $t=T/8$
(long dashes), and $t=T/4$ (short dashes). A running wave can be
discerned.

\subsection*{Fig. 7}
Plot of the ratio $U_2(t)/\overline{U}_2$ as a function of time
for swimming motion corresponding to the optimal set of
multipoles $\{\mu_{ls},\mu_{lc}\}$ with $1\leq l\leq 8$ (solid
curve). We also plot the ratio
$\mathcal{D}_2(t)/\overline{\mathcal{D}}_2$ for the same swimming
motion. This equals unity within numerical accuracy.

\subsection*{Fig. 8}
Plot of one-half the maximum eigenvalue
$\frac{1}{2}\lambda_\mathrm{max}(1,L)$ for sets of complex multipoles $\{\kappa_l,\mu_l\}$
with $1\leq l\leq L$ and $\kappa_1=0$ as a function of $L$ for $L=3,...,40$. The values tend to $\sqrt{2}$ as $L\rightarrow\infty$.

\subsection*{Fig. 9}
Plot of the reduced speed $\hat{U}^B_{1L}$ for fixed maximum
amplitude of the displacement at $\theta=\pi/2$ as a function of
$L$. At each value of $L$ the most efficient set of multipoles $\{\mu_{1s},\mu_{1c},...,\kappa_{Ls},\kappa_{Lc},\mu_{Ls},\mu_{Lc}\}$ is considered.

\subsection*{Fig. 10}
Plot of the reduced power $\hat{\mathcal{D}}^B_{1L}$ for fixed
maximum amplitude of the displacement at $\theta=\pi/2$ as a
function of $L$. At each value of $L$ the most efficient set of multipoles $\{\mu_{1s},\mu_{1c},...,\kappa_{Ls},\kappa_{Lc},\mu_{Ls},\mu_{Lc}\}$ is considered.

\subsection*{Fig. 11}
Plot of the non-vanishing components of the eigenvector with
largest eigenvalue, normalized to unity, for a set of complex multipoles $\{\kappa_l,\mu_l\}$
with $1\leq l\leq 7$ and $\kappa_1=0$. The absolute values of the $\{\kappa_l\}$ are indicated by squares and those of the $\{\mu_l\}$ are
indicated by dots.

\subsection*{Fig. 12}
Plot of the end of the displacement vector $\vc{\xi}(t)$ at
$\theta=3\pi/12,5\pi/12,\;\pi/2,7\pi/12$ and $9\pi/12$ for maximum amplitude of
the displacement at $\theta=\pi/2$ equal to $0.1\;a$ for the optimum eigenvector for $L=7$ with complex multipoles $\{\kappa_l,\mu_l\}$. The motion
is depicted with start at $t=0$ and finish at $t=\frac{7}{8}T$, where
$T=2\pi/\omega$. The endpoint is marked by a small circle.

\newpage
\setlength{\unitlength}{1cm}
\begin{figure}
 \includegraphics{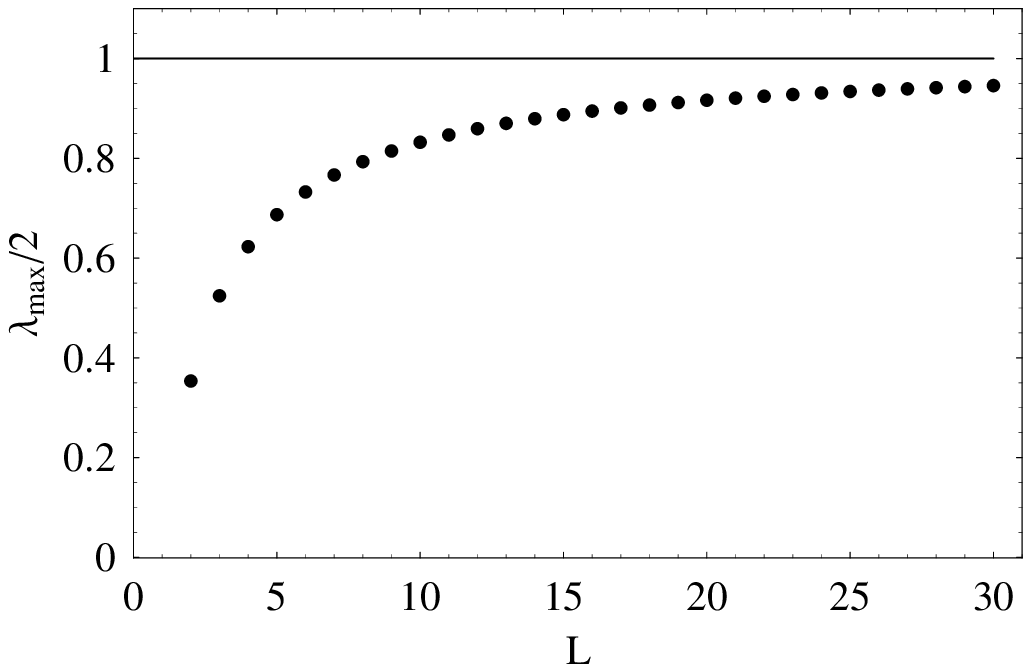}
   \put(-9.1,3.1){}
\put(-1.2,-.2){}
  \caption{}
\end{figure}
\newpage
\clearpage
\newpage
\setlength{\unitlength}{1cm}
\begin{figure}
 \includegraphics{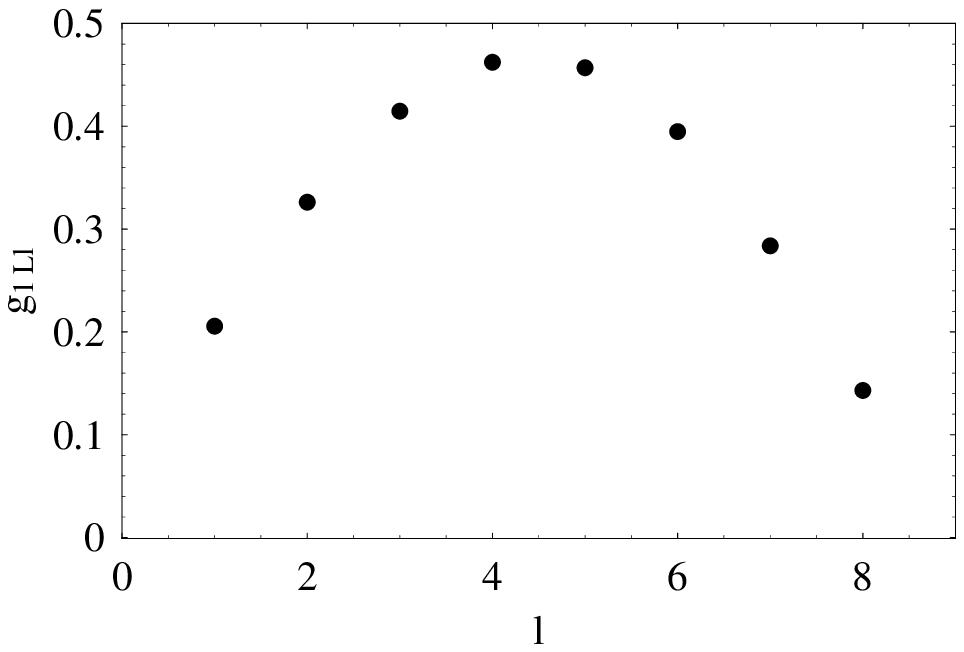}
   \put(-9.1,3.1){}
\put(-1.2,-.2){}
  \caption{}
\end{figure}
\newpage
\clearpage
\newpage
\setlength{\unitlength}{1cm}
\begin{figure}
 \includegraphics{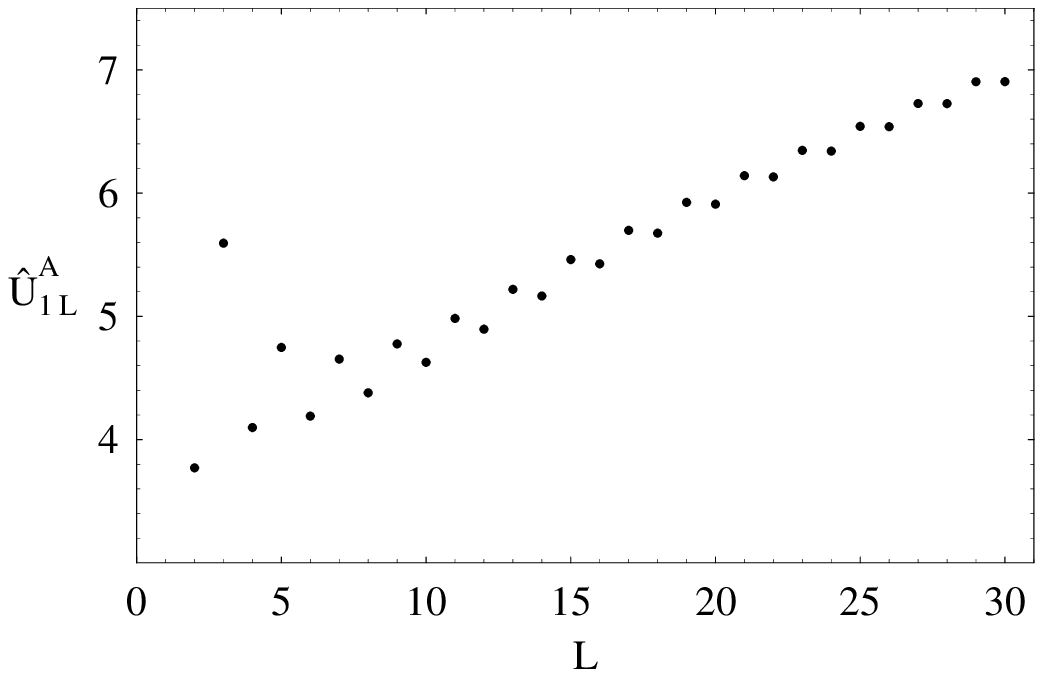}
   \put(-9.1,3.1){}
\put(-1.2,-.2){}
  \caption{}
\end{figure}
\newpage
\clearpage
\newpage
\setlength{\unitlength}{1cm}
\begin{figure}
 \includegraphics{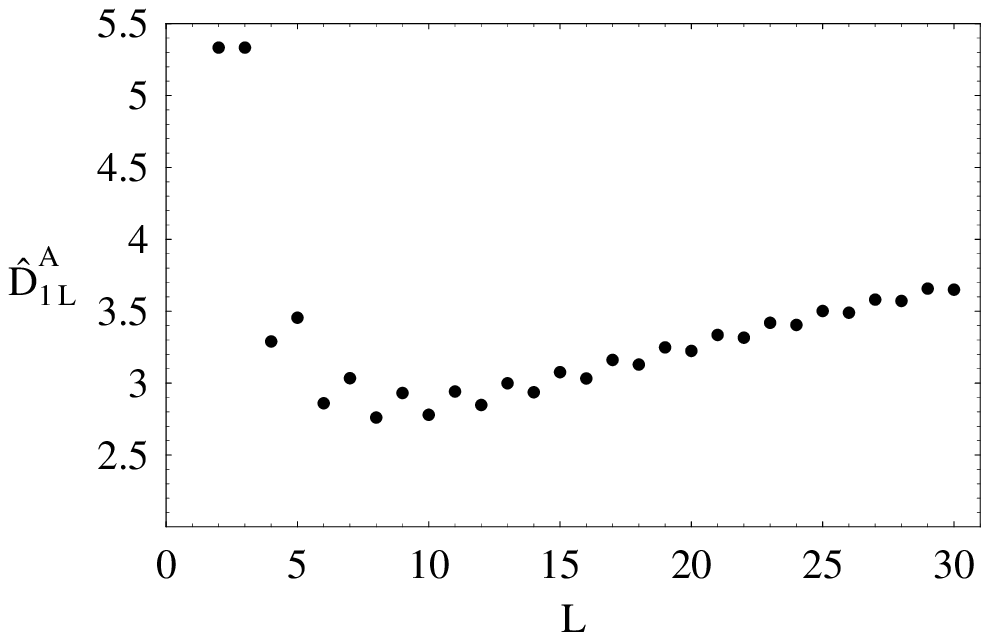}
   \put(-9.1,3.1){}
\put(-1.2,-.2){}
  \caption{}
\end{figure}
\newpage
\clearpage
\newpage
\setlength{\unitlength}{1cm}
\begin{figure}
 \includegraphics{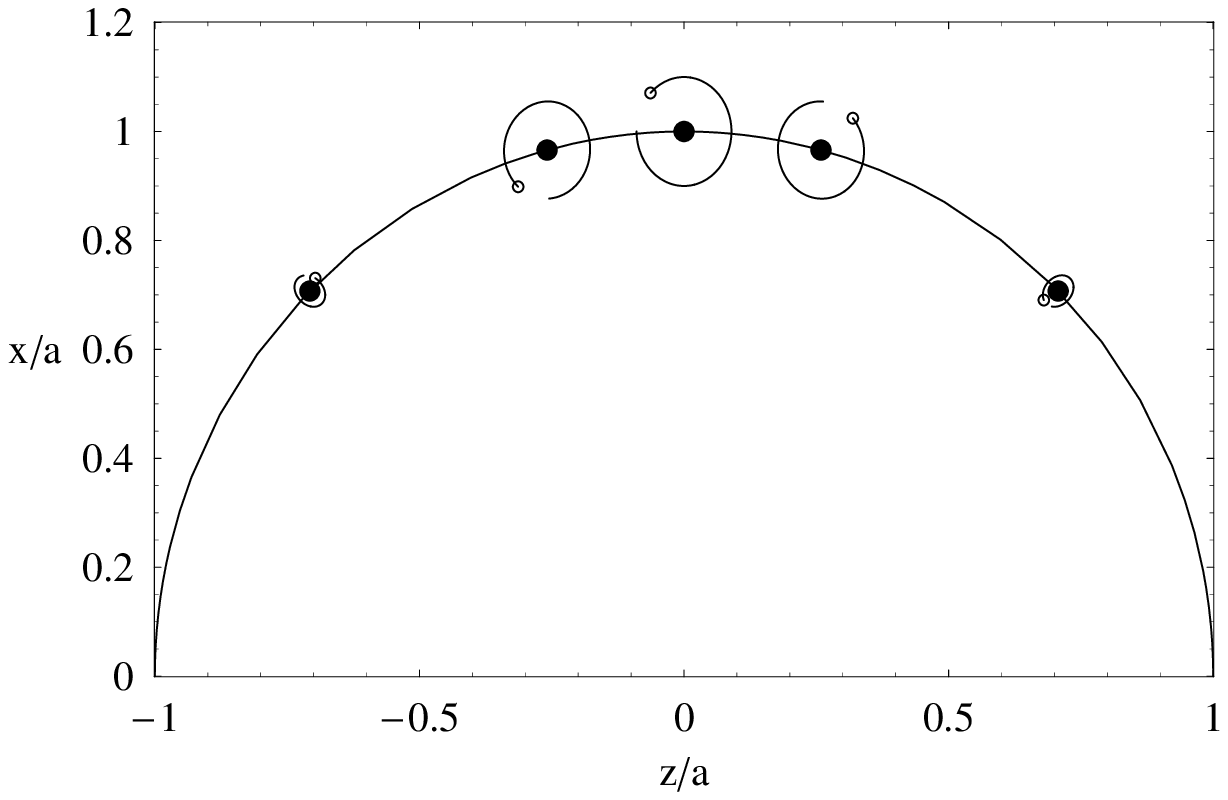}
   \put(-9.1,3.1){}
\put(-1.2,-.2){}
  \caption{}
\end{figure}
\newpage
\clearpage
\newpage
\setlength{\unitlength}{1cm}
\begin{figure}
 \includegraphics{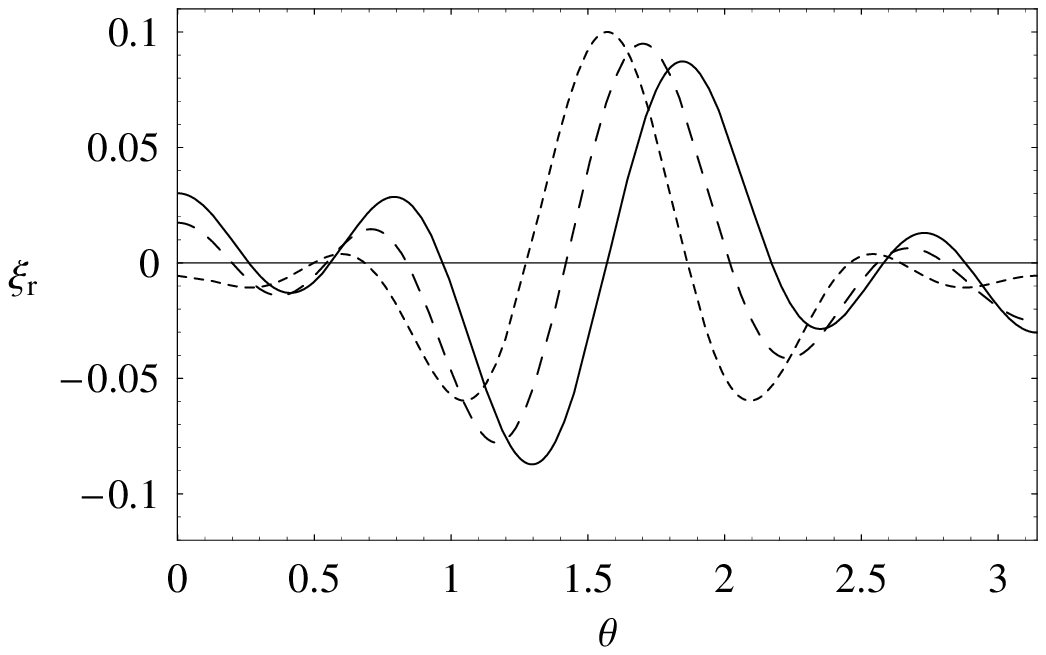}
   \put(-9.1,3.1){}
\put(-1.2,-.2){}
  \caption{}
\end{figure}
\newpage
\clearpage
\newpage
\setlength{\unitlength}{1cm}
\begin{figure}
 \includegraphics{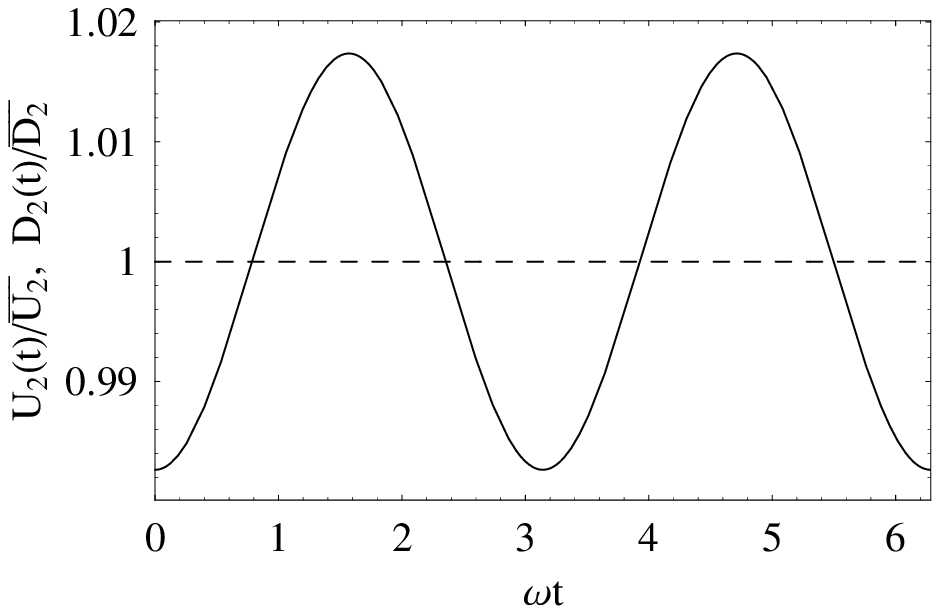}
   \put(-9.1,3.1){}
\put(-1.2,-.2){}
  \caption{}
\end{figure}
\newpage
\clearpage
\newpage
\setlength{\unitlength}{1cm}
\begin{figure}
 \includegraphics{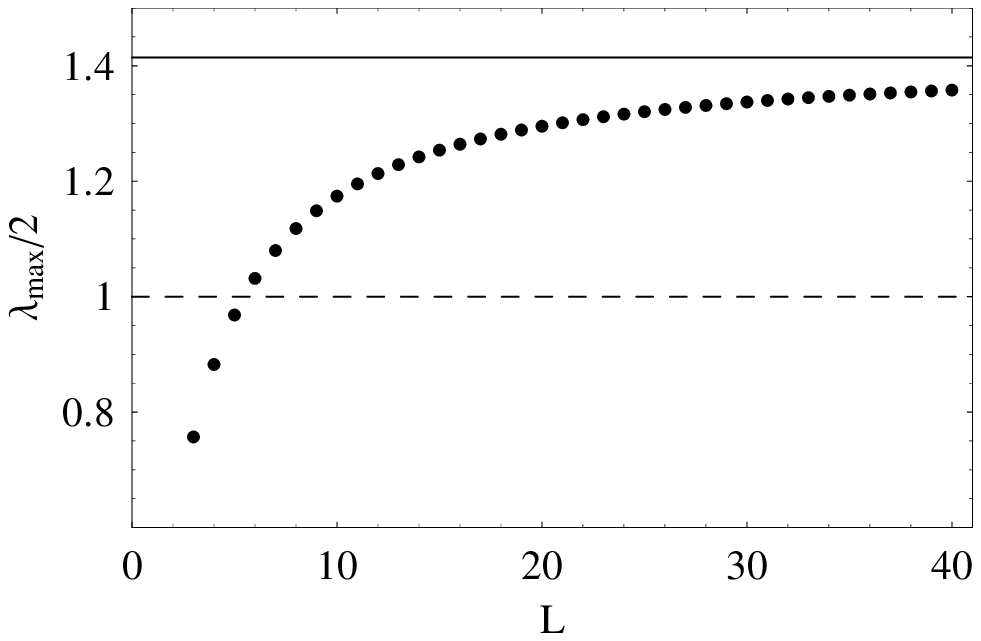}
   \put(-9.1,3.1){}
\put(-1.2,-.2){}
  \caption{}
\end{figure}
\newpage
\clearpage
\newpage
\setlength{\unitlength}{1cm}
\begin{figure}
 \includegraphics{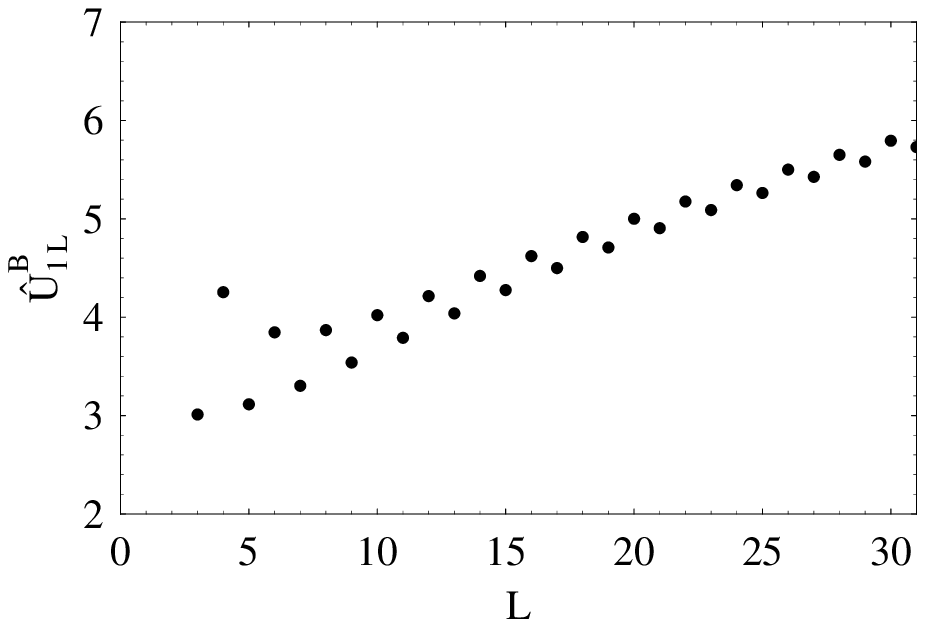}
   \put(-9.1,3.1){}
\put(-1.2,-.2){}
  \caption{}
\end{figure}
\newpage
\clearpage
\newpage
\setlength{\unitlength}{1cm}
\begin{figure}
 \includegraphics{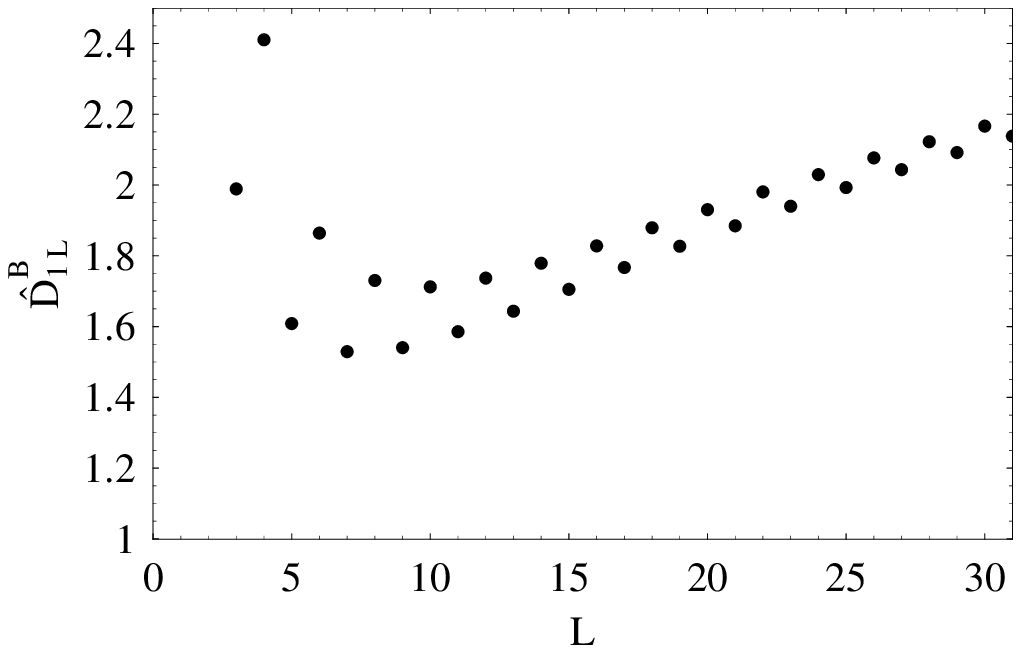}
   \put(-9.1,3.1){}
\put(-1.2,-.2){}
  \caption{}
\end{figure}
\newpage
\clearpage
\newpage
\setlength{\unitlength}{1cm}
\begin{figure}
 \includegraphics{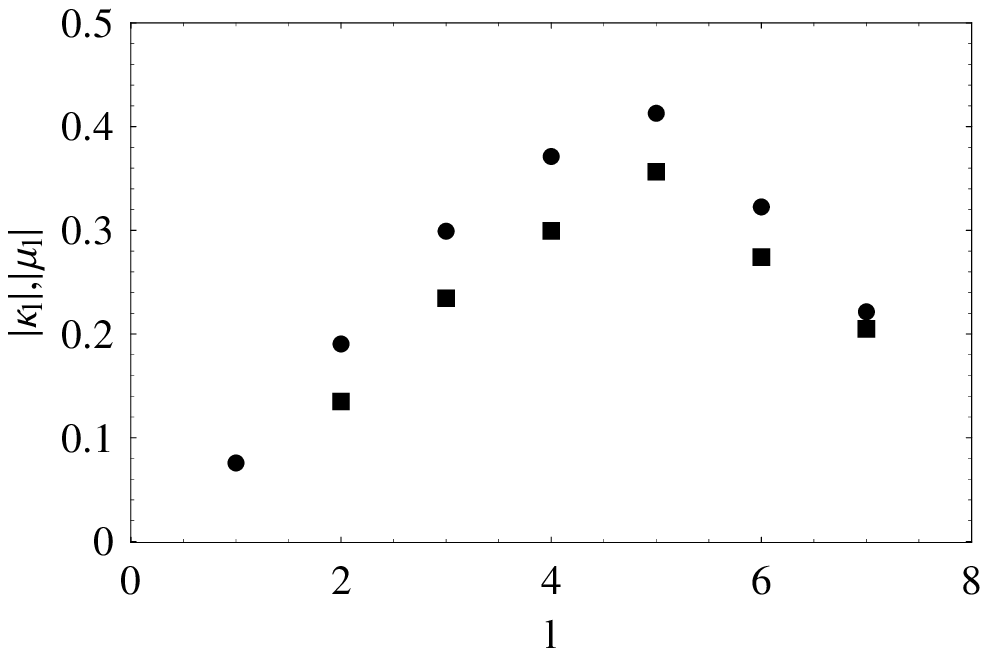}
   \put(-9.1,3.1){}
\put(-1.2,-.2){}
  \caption{}
\end{figure}
\newpage
\clearpage
\newpage
\setlength{\unitlength}{1cm}
\begin{figure}
 \includegraphics{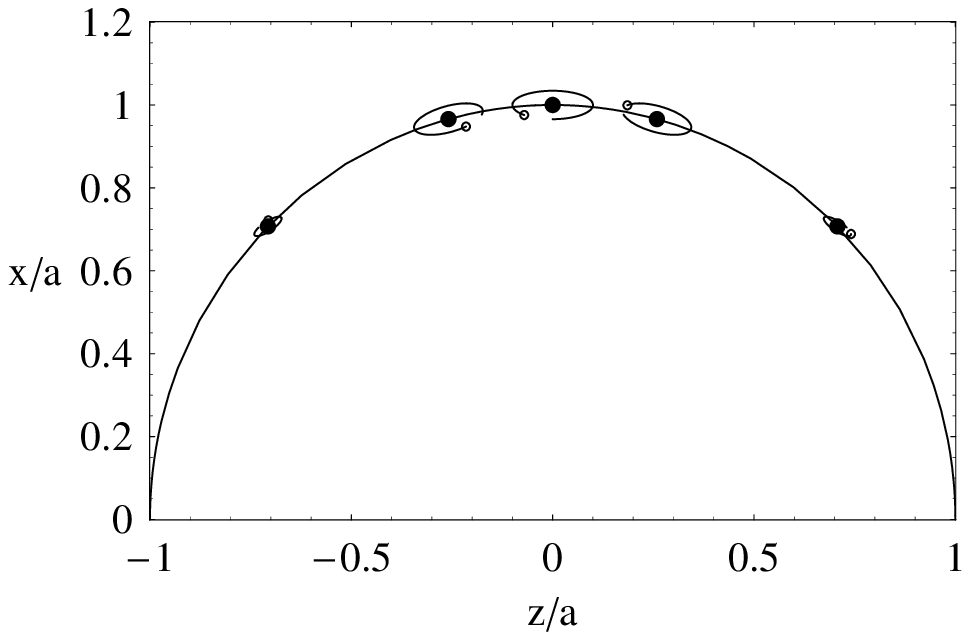}
   \put(-9.1,3.1){}
\put(-1.2,-.2){}
  \caption{}
\end{figure}
\newpage

\newpage


\begin{thebibliography}{99}

\bibitem{1}
G. I. Taylor, Proc. R. Soc. Lond. A \vol{209}, 447 (1951).

\bibitem{2}
M. J. Lighthill, Commun. Pure Appl. Maths. \vol{5}, 109 (1951).

\bibitem{3}
J. R. Blake, J. Fluid Mech. \vol{46}, 199 (1971).

\bibitem{4}
A. Shapere and F. Wilczek, J. Fluid Mech. \vol{198}, 557 (1989).

 \bibitem{5}
A. Shapere and F. Wilczek, J. Fluid Mech. \vol{198}, 587 (1989).

\bibitem{6}
J. Happel and H. Brenner, {\it Low Reynolds number hydrodynamics} (Noordhoff, Leyden, 1973).

\bibitem{7}
B. U. Felderhof and R. B. Jones, Physica A \vol{202}, 94 (1994).

\bibitem{8}
B. U. Felderhof and R. B. Jones, Physica A \vol{202}, 119 (1994).

\bibitem{9}
D. B. Dusenbery, {\it Living at Micro Scale} (Harvard University Press, Cambridge (Mass.), 2009).

\bibitem{10}
J. D. Jackson, {\it Classical Electrodynamics} (Wiley, New York, 1989).

\bibitem{11}
J. A. Sparenberg, J. Eng. Math. \vol{44}, 395 (2002).


\bibitem{12}
A. R. Edmonds, {\it Angular Momentum in Quantum Mechanics} (Princeton University Press, Princeton (N.J.), 1974).

\bibitem{13}
 M. Abramowitz and I. A. Stegun, {\it Handbook of Mathematical Functions} (Dover, New York, 1965).

\bibitem{14}
S. Childress, {\it Mechanics of swimming and flying} (Cambridge University Press, Cambridge, 1981).

\bibitem{15}
H. Tennekes, {\it The Simple Science of Flight} (MIT Press, Cambridge (Mass.), 2009).

\bibitem{16}
B. Cichocki, B. U. Felderhof, and R. Schmitz, PhysicoChem. Hyd. \vol{11}, 507 (1989).

\bibitem{17}
C. F. Bohren and D. R. Huffman, {\it Absorption and Scattering of Light by Small Particles} (Wiley, New York, 1983).

\end{thebibliography}
\end{document}